\documentclass[preprint,aps,pre,showpacs,floatfix]{revtex4}
\usepackage[dvips]{graphicx}
\usepackage{amsmath}
\usepackage{color}
\usepackage{amsfonts}
\usepackage{amssymb}
\usepackage{epstopdf}
\DeclareGraphicsExtensions{.eps,.png,.pdf}

\begin{document}

\title{Synchronous solutions and their stability in nonlocally coupled phase oscillators
       with propagation delays}
\author{Gautam C Sethia}
\email{gautam@ipr.res.in}
\author{Abhijit Sen}
\email{abhijit@ipr.res.in}
\affiliation{Institute for Plasma Research, Bhat, Gandhinagar 382 428, India}
\author{Fatihcan M. Atay}
\email{atay@member.ams.org}   
\affiliation{Max Planck Institute for Mathematics in the Sciences, Inselstr.~22-26, Leipzig D-04103, Germany}

\date{\today}


\begin{abstract}

We study the existence and stability of synchronous solutions in a continuum field of non-locally
coupled identical phase oscillators with distance-dependent propagation delays. We present a comprehensive stability 
diagram  in the parameter space of the system. From the numerical results a heuristic synchronization condition is suggested, and an analytic relation for the marginal stability curve is obtained. We also provide an expression in the form of a scaling relation that closely follows the marginal stability curve over the complete range of the non-locality parameter. 
\end{abstract}
\maketitle

\section{Introduction}
Phase oscillators are simple models of oscillatory activity where the
essential variable is an angular position along a limit cycle. Systems of
coupled phase oscillators provide a convenient paradigm for studying the
collective behavior of weakly interacting limit cycle oscillators
\cite{winfree67,Kuramoto,ermentrout84} 
and are in turn useful for modeling a wide variety 
of cooperative phenomena
observed in biological, physical and chemical systems \cite{walker69,buck88,peskin75,strogatz92,wiesenfield96}.
One of the most well known models of coupled phase oscillators is the Kuramoto
model \cite{Kuramoto} which considers a system of globally
(all-to-all) coupled set of oscillators with a spread of intrinsic frequencies. 
The oscillators synchronize to a common frequency as soon as the coupling strength 
goes beyond a threshold value. For a system of identical oscillators one achieves phase 
synchrony where, starting from an initial 
random distribution of phases, all the oscillators acquire the same phase within a time 
that is inversely proportional to the coupling strength. When the members of the set of 
identical oscillators have a
spatial identity (e.g. when the coupling is of the nearest-neighbour kind or varies 
in strength over space) the collective states can exhibit a broader class of phase-locking
behaviour including acquiring the same phase (synchrony) or a fixed amount of phase difference between 
neighbouring oscillators (traveling waves). For a nonlocally coupled continuous system of this kind it
is also possible to sustain a more exotic state whereby phase-locked and incoherent activity 
can simultaneously exist at different spatial locations \cite{kuramoto02}, 
giving rise to a
spatio-temporal pattern that has been termed as a \emph{chimera state }
\cite{abrams04}. 

While the original Kuramoto model is based on the idealized assumption of instantaneous coupling between the oscillators, there has been a growing interest in extending the model to include finite time delay in the coupling and to study the existence
and stability of phase-locked states in such a situation \cite{schuster89,niebur91,kim97,crook97,yeung99,zanette00,earl03}. Time
delayed coupling is physically relevant in many real life systems and accounts for finite propagation times of signals, reaction times of chemicals, individual neuron firing periods in neural networks
etc. One of the earliest investigations in this direction was made by 
Schuster and Wagner \cite{schuster89} who considered the collective states of a pair of delay-coupled phase oscillators
and observed changes in the onset conditions for synchrony as well as the existence of additional (higher) branches 
of synchronized frequency states. Niebur et al.~\cite{niebur91} considered a two-dimensional array of phase
oscillators that interact via time delayed nearest neighbour coupling and found that time delay led to 
a reduction in the collective frequency of the system. Their numerical studies also found higher frequency states which were metastable and decayed to the lowest branch in the presence of noise. Kim et al \cite{kim97} studied multistability of synchronized and desynchronized states in a time-delayed Kuramoto system in the presence of a pinning force. More recently Yeung and Strogatz \cite{yeung99} have carried out a systematic stability study of the synchronous states of a delayed Kuramoto system and have obtained analytic stability curves for the limiting case of coupled identical oscillators. Since the original Kuramoto model adopts a global coupling between the oscillators, its time-delayed version considered in all the above mentioned studies have assumed a single fixed amount of delay in the coupling. In a more realistic situation where the oscillators are spatially distributed and are coupled in a nonlocal fashion, such an assumption is not appropriate and one needs to consider distance-dependent delays. 

Models with distance-dependent delays have been proposed in the past and analyzed under simplifying assumptions or in convenient limits. Crook et al.~\cite{crook97} considered a continuum system of coupled identical oscillators with a spatially decaying interaction kernel and modeled the space dependent time delay contribution through an effective phase shift term in the interaction. They also took the size of the system to be infinite for mathematical convenience but thereby effectively reduced the nature of the mutual coupling to a ``local'' one (since the interaction length is always much smaller than the system length). Zanette \cite{zanette00} studied another simplified version of the generalized system where he adopted a {distance} {independent} (global) coupling between the oscillators but introduced a {distance} {dependent} time delay in the interaction. He obtained numerical results on the stability of synchronous and propagating traveling waves and also some analytic results in the limit of small delay. Ko \& Ermentrout \cite{ko07} recently investigated the effects of distance dependent delays in sparsely connected oscillator systems and found that a small fraction of connections with time delay can destabilize the synchronous states. We note that distance-dependent delays have also been considered in other related contexts, e.g., in the stability of equilibria of continuum models of neural tissue with axonal propagation delays \cite{PhysD05,SIADS06
}.  
To the best of our knowledge, however, there has been no systematic study of the synchronization properties of the full nonlocal system of phase oscillators with distance-dependent delays. Our present work is motivated by a desire to address this problem. The presence of both non-locality and time delay in the coupling pose a formidable mathematical challenge for an analytic solution of the problem and one has to perforce adopt a numerical approach. In the recent past we have studied the generalized system to obtain {\it chimera states} in the presence of delay \cite{sethia08}. In the present work we carry out a detailed study on the existence and stability properties of synchronous solutions of the system. Through detailed stability curves in the parameter space of the mean time delay ($\bar{\tau}$) and the non-locality parameter ($\kappa$) (for a fixed intrinsic oscillator frequency $\omega$) we delineate the relative influence of  $\bar{\tau}$ and $\kappa$ on the stability of the synchronous state whose frequency is $\Omega$. 
 From an extensive scan over several frequencies $\omega$, we further obtain a comprehensive stability diagram in the parameter space of $\Omega \bar{\tau}$ and $\kappa$ that is independent of $\omega$. Our numerical results further suggest that the instability mechanism corresponds to a saddle-node bifurcation and the most unstable perturbation has the lowest mode number. These results help us obtain an analytic expression for the synchronization condition.  
The synchronization condition
can be closely approximated by an offset exponential power law relationship between $\Omega \bar{\tau}$ and $\kappa$, which is valid over a wide range of coupling regimes and provides a useful criterion for determining the critical value of the delay-induced maximum phase shift below which the synchronous state is stable. 

\section{Model system and its synchronous states}
We consider a continuum of identical phase oscillators, arranged on a circular ring $C$
and labeled by $x\in\lbrack-L,L]$ with periodic boundary conditions, whose
dynamics is governed by
\begin{equation}
\frac{\partial}{\partial t}\phi(x,t)  =\omega-K\int_{-L}^{L
}G(|z|)\sin\left[  \phi(x,t)-\phi\left( x-z,t-\frac{|z|}{v}\right)  \right]dz, 
\label{pha}
\end{equation}
where $\phi(x,t) \in [0,2\pi)$ is the phase of the oscillator at location $x$ and time $t$, whose intrinsic oscillation frequency is $\omega>0$, $K$ is the coupling strength and $G:[-L,L]\rightarrow\mathbb{R}$ is an even function
describing the coupling kernel.  The quantity $v$ denotes the signal
propagation speed which gives rise to a time delay of $|z|/v$ for distance $|z|$ from  the locations $x$. As the oscillators are located on a ring with circumference $2L$, the distance between any two oscillators is given by the shorter of the Euclidean distance between them along the ring. In this configuration, the maximum distance between the coupled oscillators is $L$ and thus the maximum time delay would be $\tau_m=L/v$. 

We choose the coupling kernel $G(|z|)$ to have an exponentially decaying nature and its normalized form is taken to be 
\begin{equation}
G(|z|)=\dfrac{\sigma}{2(1.0-e^{-L \sigma})}e^{-\sigma|z|}, \label{ker} 
\end{equation}  
where $\sigma > 0$ is the inverse of the interaction scale length and is a measure of the nonlocality of the coupling.
We make time and space dimensionless in  Eqs.(\ref{pha}) and (\ref{ker}) by the transformations $t\rightarrow Kt$, $\omega\rightarrow\omega/K$,
$\kappa \rightarrow \sigma L$, $z \rightarrow z / L$, $\tau_m \rightarrow K\tau_m $ and $x \rightarrow x / L$   and obtain

\begin{equation}
\frac{\partial}{\partial t}\phi(x,t)  =\omega-\int_{-1}^{1
}G(|z|)\sin\left[  \phi(x,t)-\phi\left( x-z,t-|z|\tau_{m}\right) \right]dz. 
\label{phase}
\end{equation}

\begin{equation}
G(|z|)=\dfrac{\kappa}{2(1.0-e^{-\kappa})}e^{-\kappa|z|} \label{kernel} 
\end{equation}

We look for synchronous solutions of Eq.~(\ref{phase}) that have the form:
\begin{equation}
\phi_{\Omega}(t)=\Omega t+\phi_{0}. \label{sol}
\end{equation}
The value of $\phi_{0}$ can be taken to be zero by a translation. Substituting Eq.~(\ref{sol}) in Eq.~(\ref{phase})
we get

\begin{align}
\Omega &=\omega-\int_{-1}^{1}G(|z|)\sin\left(  \Omega\tau_{m}|z|
\right)   \,dz \nonumber\\
 &=\omega-\frac{\kappa}{(1-e^{-\kappa})}\left[ \frac{\Omega \tau_{m}-\Omega \tau_{m}e^{-\kappa}\cos(\Omega \tau_{m})-\kappa e^{-\kappa} \sin(\Omega \tau_{m})}{\kappa^{2}+(\Omega\tau_{m})^{2}}\right], \label{eqlbm}
\end{align}
which is an implicit equation in $\Omega$. Being a transcendental equation its solution can in principle be multi-valued in $\Omega$ for a given set of parameters $\omega, \tau_{m}$ and $\kappa$, and can lead to higher branches of collective frequencies as pointed out by Schuster and Wagner \cite{schuster89} for a system of two coupled oscillators. 

We further define a mean delay parameter by
\begin{equation}
\bar{\tau}=\int_{-1}^{1}G(|z|)\tau_{m}|z|\, dz\
\label{mtaueqn}
\end{equation}
which weights the individual delays with the corresponding connection
weights. With the exponential connectivity given by Eq.~(\ref{kernel}), this translates
into
\begin{equation}
\bar{\tau}=\tau_{m}\frac{e^{\kappa}-\kappa-1}{\kappa(e^{\kappa}-1)}.
\label{mtau}
\end{equation}
This gives values for the limiting cases: $\bar{\tau}\to0$
for local ($\kappa\to \infty$) and $\bar{\tau}\to\tau_{m}/2$ for global ($\kappa\to 0$) coupling.


Fig.~1 plots the numerical solutions $\Omega$ of Eq.~(\ref{eqlbm})  as a function of $\bar{\tau}$ for $\kappa=2.0$ and several values of $\omega$. 
As the curve for $\omega=0.8$ shows, it is possible to have multiple solutions $\Omega$ for a given value of $\bar{\tau}$. The stability of these higher states will be discussed in later sections of the paper. One also notes that the lowest branch shows frequency suppression as a function of the mean time delay $\bar{\tau}$.

We can also examine two simple limits of Eq.~(\ref{eqlbm}) which help us  understand the influence of the non-locality parameter $\kappa$ on the nature of the equilibrium solutions. For $\kappa \rightarrow 0$,  corresponding to the global coupling limit, Eq.~(\ref{eqlbm}) reduces to,
\begin{equation}
\Omega = \omega-\frac{1-\cos(\Omega \tau_{m})}{\Omega\tau_{m}} \label{eqlbm_z}
\end{equation}
This was the limiting case examined by Zanette \cite{zanette00}, where the transcendental nature of Eq.~(\ref{eqlbm}) was retained, thereby permitting several multiple branches of solutions to exist, but the coupling was taken to be global. In the limit of $\Omega \tau_m \ll 1$, (\ref{eqlbm_z}) gives,
\begin{equation}
\Omega \approx \frac{\omega}{1 + \tau_m/2}
\end{equation}
demonstrating frequency suppression of the collective mode due to finite time delay in this parameter regime.
The other interesting scenario is that of $\kappa \gg 1$, which corresponds to the situation of local coupling among the oscillators. In this case Eq.~(\ref{eqlbm}) becomes
\begin{equation}
\Omega = \omega - \frac{\kappa \Omega \tau_m}{\kappa^2 + \Omega^2\tau_m^2}\label{eqlbm_local}
\end{equation}
Eq.~(\ref{eqlbm_local}) is a third order polynomial equation in $\Omega$ and can at most have three real solutions in contrast to the higher number of multiple roots of the transcendental Eq.~(\ref{eqlbm}). 
The local limit can also be approached by taking $L \rightarrow \infty$ limit in the
unscaled form of Eq.~(\ref{eqlbm}) in which  $\tau_m$ is replaced by $L/v$ and $\kappa$ by $\sigma L$. The mean delay ($\bar{\tau}_{\infty}$) in this system equals $\frac{1}{\sigma v}$ and Eq.~(\ref{eqlbm}) can be rewritten in terms of $\bar{\tau}_{\infty} $ as

\begin{equation}
\Omega = \omega - \frac{ \Omega \bar{\tau}_{\infty}}{1 + \Omega^2 {\bar{\tau}_{\infty}}^2}\label{Omega_inf}
\end{equation}
Crook {\it et al.}~\cite{crook97} considered such a limit but with an additional simplifying
assumption regarding the contribution of the delay term: Instead of explicitly treating the delay term in the
argument of the phase function they modeled the delay contribution by a space dependent phase shift (implicit delay). Thus the
basic model Eq.~(\ref{pha}) was simplified in their case to be of the form
\begin{equation}
\frac{\partial}{\partial t}\phi(x,t)  =\omega-K\int_{-\infty}^{\infty
}G(|z|)\sin\left[  \phi(x,t)-\phi( x-z,t)+\frac{|z|}{v}\right]dz 
\label{pha_crook}
\end{equation}
with $K=1$.
The corresponding equation for the equilibrium solutions is given in this case by
\begin{equation}
\Omega = \omega - \frac{\bar{\tau}_{\infty}}{1 + {\bar{\tau}_{\infty}}^2}\label{crook}
\end{equation}
which shows only a single collective frequency that experiences time delay induced reduction for  $\bar{\tau}_{\infty}<1$ and a slow rise to $\omega$ for $\bar{\tau}_{\infty} > 1$. 

The degree to which the non-locality parameter $\kappa$ can influence the frequency of the synchronous equilibrium solutions of the coupled oscillator system can be seen in Fig.~\ref{fig:fig2}. This figure provides a graphical comparison of this influence by displaying plots of $\Omega$ {\it vs} $\bar{\tau}$ obtained by solving Eq.~(\ref{eqlbm}) with different values of $\kappa$. 
The curve labeled  \textit{NL} represents a typical case of nonlocal coupling, here calculated for $\kappa=2.0$. The curve \textit{L} is for local coupling with $\kappa =10.0$, whereas the global coupling curve \textit{G} is calculated for $\kappa=0.01$. The $G$ and $L$ curves obtained from Eq.~(\ref{eqlbm_z}) and Eq.~(\ref{eqlbm_local}) respectively are not significantly different from what are shown in this figure. The curve $L_{\infty}$ plots Eq.~(\ref{crook}) for comparison. We have used $\omega=0.8$ for the plots shown in this figure. 

\section{Stability of the synchronous solutions}

We now examine the stability of the synchronous solutions $\phi_{\Omega}$ obtained in the previous section. 
\subsection{Eigenvalue equation}
The linear stability of solutions $\phi_{\Omega}(t)$ of Eq.~(\ref{phase})
is determined by the variational equation
\begin{equation}
\frac{\partial}{\partial t}u(x,t)  = -\int_{-1}^{1}G(|z|)\cos\left[\Omega\tau_{m} |z| \right][u(x,t)-u(x-z,t-|z|\tau_{m})]dz
\label{veqn}
\end{equation}
where $u(x,t)=\phi(x,t)-\phi_{\Omega}(t)$. With the ansatz
$u(x,t) \sim e^{\lambda t}e^{i\pi nx}$, $\lambda\in\mathbb{C}$, $n\in\mathbb{Z}$, we
obtain the eigenvalue equation
\begin{equation}
\lambda =-\int_{-1}^{1}G(|z|)\cos\left(  \Omega\tau_{m}|z|\right) 
\left(  1-e^{-\lambda|z|\tau_{m}}e^{-i\pi n z}\right)  \,dz.
\label{delay-lambda}
\end{equation}
Writing $\lambda = \lambda_R + i\lambda_I$ and separating Eq.~(\ref{delay-lambda}) into its real and imaginary parts we get
\begin{align}
\lambda_R & = -\int_{-1}^{1}G(|z|)\cos\left(  \Omega |z|\tau_{m}\right)\left[  1-e^{-\lambda_R|z|\tau_{m}}\cos\left( \lambda_I |z|\tau_{m} + \pi nz\right)\right]dz,\label{lambda_R}\\
\lambda_I & = -\int_{-1}^{1}G(|z|)\cos\left(  \Omega |z|\tau_{m}\right)e^{-\lambda_R|z|\tau_{m}}\sin\left(\lambda_I|z|\tau_{m} + \pi nz\right)dz.\label{lambda_I}
\end{align}
Since the perturbations $u(x,t)$ corresponding to $n=0$ again yield synchronous solutions, 
the linear stability of the synchronous state requires that all solutions of Eq.~(\ref{delay-lambda}) have $\lambda_R < 0$ for all non-zero integer values of $n$. The marginal stability curve in the parameter space of $(\kappa,\tau_m)$ is defined by $\lambda_R=0$ and in principle can be obtained by setting $\lambda_R=0$ in Eq.~(\ref{lambda_R}), solving it for $\lambda_I$ and 
substituting it in Eq.~(\ref{lambda_I}). In practice it is not possible to carry out such a procedure analytically for the integral Eqs.~$(\ref{lambda_R},\ref{lambda_I})$ and one needs to adopt a numerical approach, which is discussed in  the next section. 

\subsection{Numerical determination of eigenvalues}

To systematically determine the eigenvalues of Eq.~(\ref{delay-lambda}) in a given region of the complex plane we use multiple methods in a complementary fashion. 
First, Eq.~(\ref{eqlbm}) is solved for
$\Omega$ for a given set of values of the parameters $\kappa$, $\omega$ and $\tau_{m}$. Following that, we need to determine the complex zeros of the function $f(\lambda)$ defined as
$$ f(\lambda) = \lambda + \int_{-1}^{1}G(|z|)\cos\left(  \Omega\tau_{m}|z|\right) 
\left(  1-e^{-\lambda|z|\tau_{m}}e^{-i\pi n z}\right)  \,dz,$$
which is equivalent to finding solutions $\lambda$ of (\ref{delay-lambda}).
To do this we have primarily relied on the numerical technique developed by Delves and Lyness \cite{delves67} based on the Cauchy's argument principle. By this principle the number of unstable roots $m$ of $f(\lambda)$ is given by
$$ m = \frac{1}{2\pi i}\oint_C\frac{f'(\lambda)}{f(\lambda)}d\lambda, $$
where the closed contour $C$ encloses a domain in the right half of the complex $\lambda$ plane with 
the imaginary axis forming its left boundary. Once we get a finite number for $m$ we further trace the location of the roots by plotting  the zero value contour lines of the real and imaginary parts of the function $f(\lambda)$ in a finite region of the complex plane $(\lambda_R,\lambda_I)$. 
The intersections of the two sets of contours locate all the eigenvalues of Eq.~(\ref{delay-lambda}) in the given region of the complex plane. The computations are done on a fine enough grid (typically $80 \times 80$) to get a good resolution. A systematic scan for unstable roots is made by repeating the above procedure for many values of the perturbation number $n$ and by gradually extending the region of the complex plane. We have made extensive use of {\it Mathematica} in obtaining the numerical results on the stability of the synchronous states.

\subsection{Results}

\begin{figure}
\includegraphics[width=15cm]{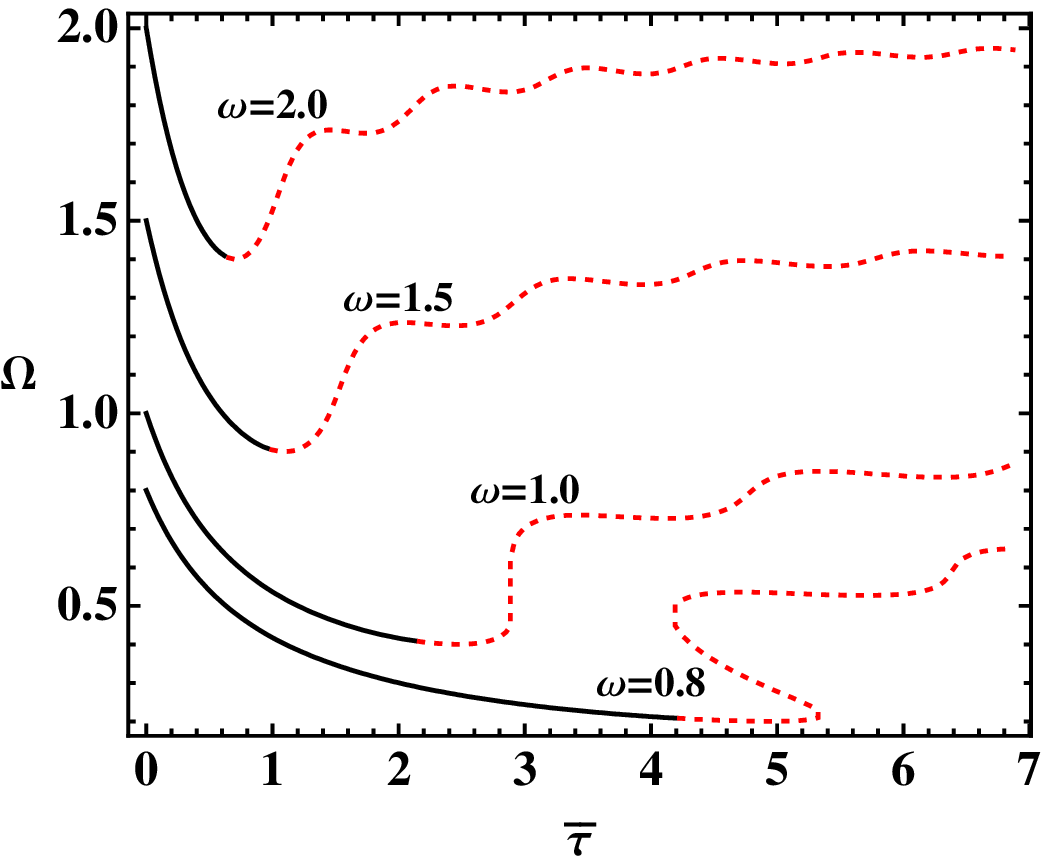}
\vskip.0cm
\caption{Frequency $\Omega$ of the synchronous solutions of Eq.~(\ref{phase}) for $\kappa=2.0$ as a function of the mean time delay $\bar{\tau}$. The different curves correspond to solutions of Eq.~(\ref{eqlbm}) for different values of the intrinsic oscillator frequency $\omega$. The solid portions denote stable states and the dotted ones (in red) unstable states.}
\label{fig:fig1}
\end{figure}

\begin{figure}
\includegraphics[width=15cm]{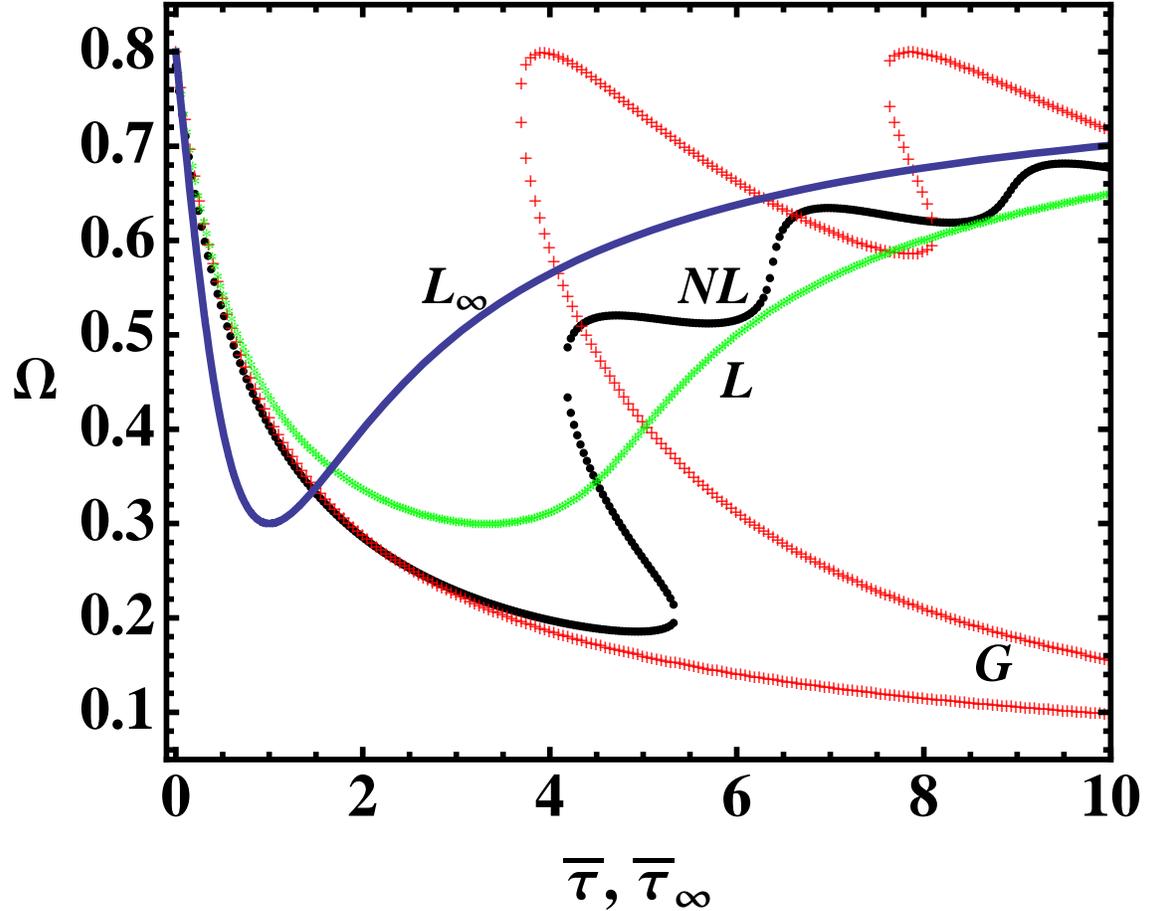}
\vskip.0cm
\caption{The frequency of synchronous oscillations (\textit{G, L} and \textit{NL})  obtained from Eq.~(\ref{eqlbm}) for a fixed value of $\omega=0.8$ and for different values of $\kappa$, representing global coupling $G$ ($\kappa=0.01$), local coupling $L$ ($\kappa=10.0$), and intermediate (non-local) coupling $NL$ ($\kappa=2.0$). The curve $L_{\infty}$ has been obtained from Eq.~(\ref{crook}).}
\label{fig:fig2}
\end{figure}

\begin{figure}
\includegraphics[width=15cm]{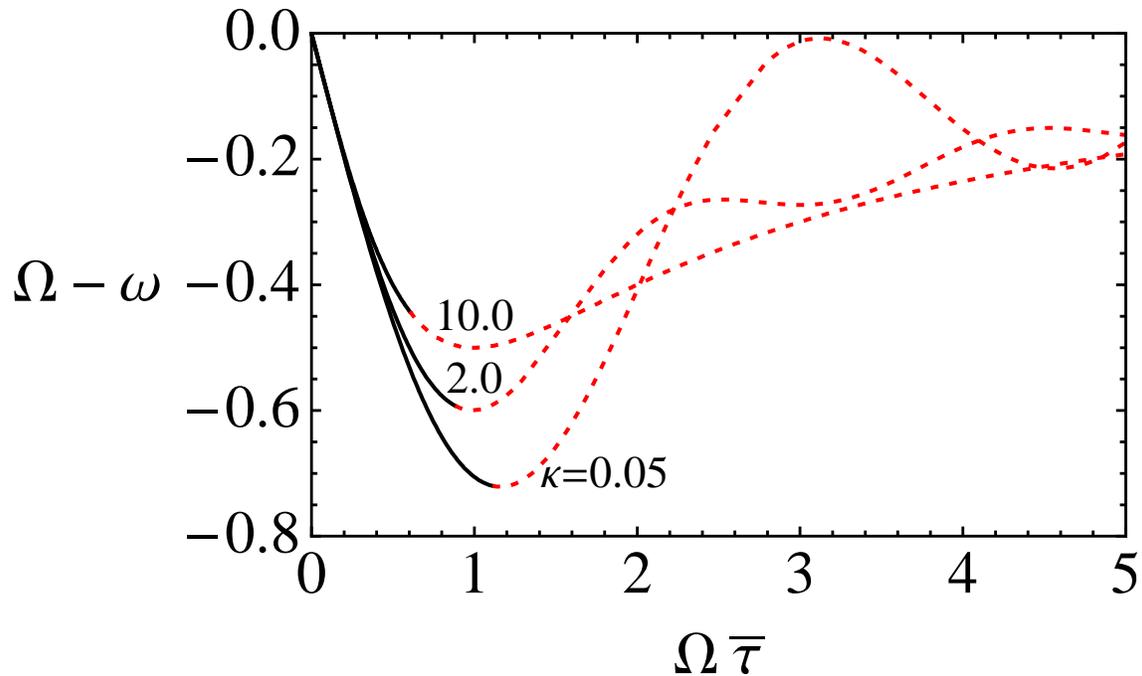}
\vskip.0cm
\caption{Solutions of Eq.~(\ref{eqlbm}) for the synchronous oscillation frequency for several values of $\kappa$, plotted in terms of $\Omega - \omega$ versus $\Omega \bar{\tau}$. Note that $\Omega-\omega=H(\Omega \bar{\tau},\kappa)$ for the equilibrium solutions (see Eq.(\ref{H})). In this representation, the different curves of Fig.1 corresponding to different values of $\omega$  collapse onto a single curve for a given value of $\kappa$. The solid (black) portions of the curves correspond to stable synchronous states and the dotted ones (red) to unstable synchronous states.}
\label{fig:fig3}
\end{figure}

In Fig.~\ref{fig:fig1} the solid portion of the curve shows the stable synchronous states of 
Eq.~(\ref{phase}) for $\kappa=2.0$ and for various values of $\omega$ and $\bar{\tau}$. The terminal point on a given  solid curve of $\Omega$ {\it vs} $\bar{\tau}$ marks the marginal stability point. The marginal stability point is seen to shift towards larger values of $\bar{\tau}$  as one moves down to curves with lower
values of $\omega$. 
A more compact representation is obtained if one plots 
$\Omega-\omega$ versus $\Omega\bar{\tau}$, since in this case the solutions corresponding to different values of $\omega$ for a given $\kappa$ consolidate onto a single curve,
as shown in Fig.~\ref{fig:fig3} for $\kappa= 0.05,2.0$ and $10.0$ respectively. 

It is seen from both figures that the stability domains of the synchronous solutions are restricted to the lowest branch where the curves are decreasing.
This suggests a heuristic necessary (but not sufficient) condition for stability of the synchronous solutions: From Fig.~\ref{fig:fig1} we have that 
$\partial\Omega / \partial \bar\tau < 0$ for stable synchronous solutions, and
from Eq.~(\ref{eqlbm}) we calculate
\begin{equation}  \label{d-omega}
	\frac{\partial\Omega}{\partial \bar{\tau}} = 
	\frac{-\Omega c_\kappa I} {1 + c_\kappa\bar{\tau} I}
\end{equation}
where
\begin{equation}  \label{I}
	I =  \int_{-1}^{1} |z| G(|z|) \cos(c_\kappa \Omega \bar\tau |z|) \,dz
\end{equation}
and
\[
	c_\kappa = \frac{\kappa(e^\kappa - 1)}{e^\kappa - 1 - \kappa}.
\]
Since $c_\kappa > 0$ and $I$ is bounded, the denominator in Eq.~(\ref{d-omega}) is positive for small values of $\bar{\tau}$. Hence, for positive $\Omega$, the requirement $\partial\Omega / \partial \bar\tau < 0$ implies the condition $I>0$, 
that is,
\begin{equation}
\int_{-1}^{1}|z|G(|z|)\cos\left( c_\kappa \Omega \bar\tau |z|\right)dz\;>\;0
\label{nec_cond}
\end{equation} 

An alternative approach to arrive at the necessary condition (Eq.\ref{nec_cond}) would be to make use of the results presented in Fig.\ref{fig:fig3}. We recast the dispersion relation given by  Eq.(\ref{eqlbm}) in the form:
\begin{equation}
 \Omega-\omega = H(\Omega \bar{\tau},\kappa)
\label{H}
\end{equation} 
where
$$H(\Omega \bar{\tau},\kappa) = -\int_{-1}^{1}G(|z|)\sin\left( c_\kappa \Omega \bar\tau  |z|\right) dz $$ 
It is seen from Fig.\ref{fig:fig3} that the stability domain of the synchronous solutions is restricted to the lowest branch where the curves have a negative slope. This again suggests a heuristic necessary condition for stability of the synchronous solutions to be $H'<0$ leading to the necessary condition given by Eq.(\ref{nec_cond}). The prime indicates a derivative of $H(\Omega \bar{\tau},\kappa)$ $w.r.t$ $\Omega \bar{\tau}$. 

As we will see later, the marginal stability curve obtained from $H'$ or $ I = 0$ does lie above the true marginal stability curve (see Fig.~\ref{fig:fig5}), confirming that
condition (\ref{nec_cond}) is necessary but not sufficient for the stability of synchronous states.
 
\begin{figure}
\begin{tabular}[c]{cc}
\includegraphics[width=15cm]{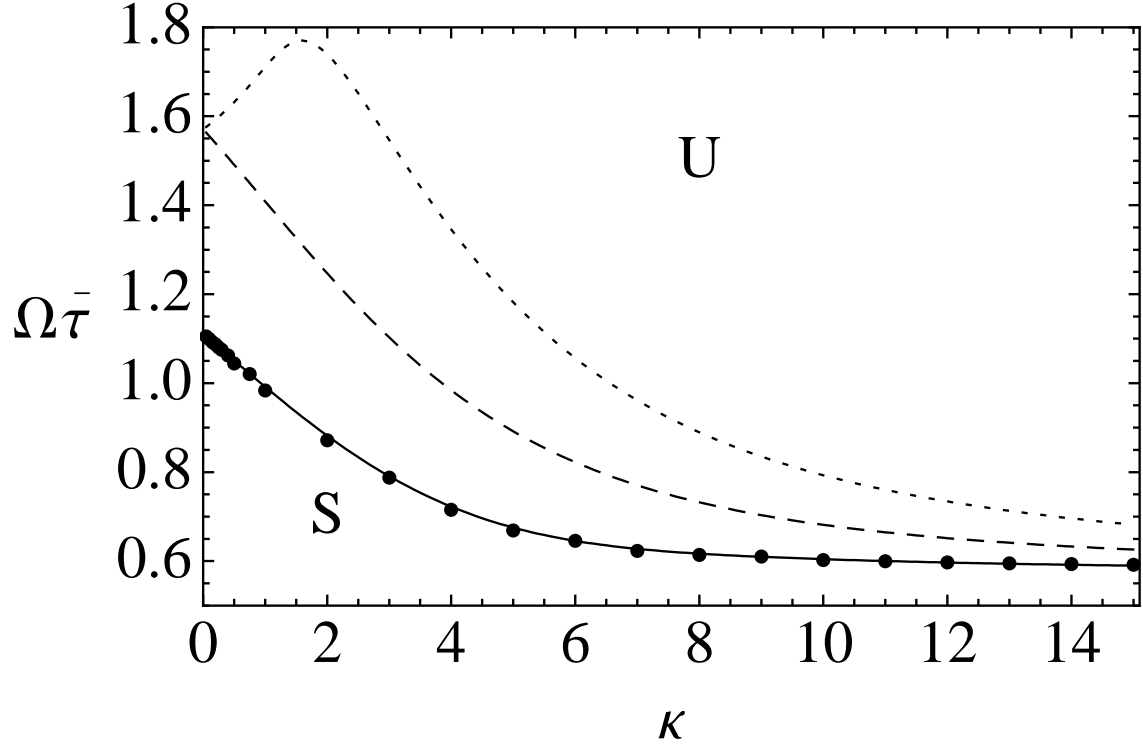}
\end{tabular}
\caption{The marginal stability curve (solid curve) in the ($\Omega\bar{\tau},\kappa$) space, obtained from the lowest branch solutions of Eq.~(\ref{marg_stab}) for $n=1$. The filled
circles correspond to numerical results from eigenvalue analysis of Eq.~(\ref{delay-lambda}), and show a perfect fit to the analytical result. The dashed and dotted curves correspond to marginal stability curves
obtained for $n=2$ and $n=3$ perturbations, respectively. The symbols $S$ and $U$ denote stable and unstable regions in the parameter space.}
\label{fig:fig4}
\end{figure}

\begin{figure}
\begin{tabular}[c]{cc}
\includegraphics[width=15cm]{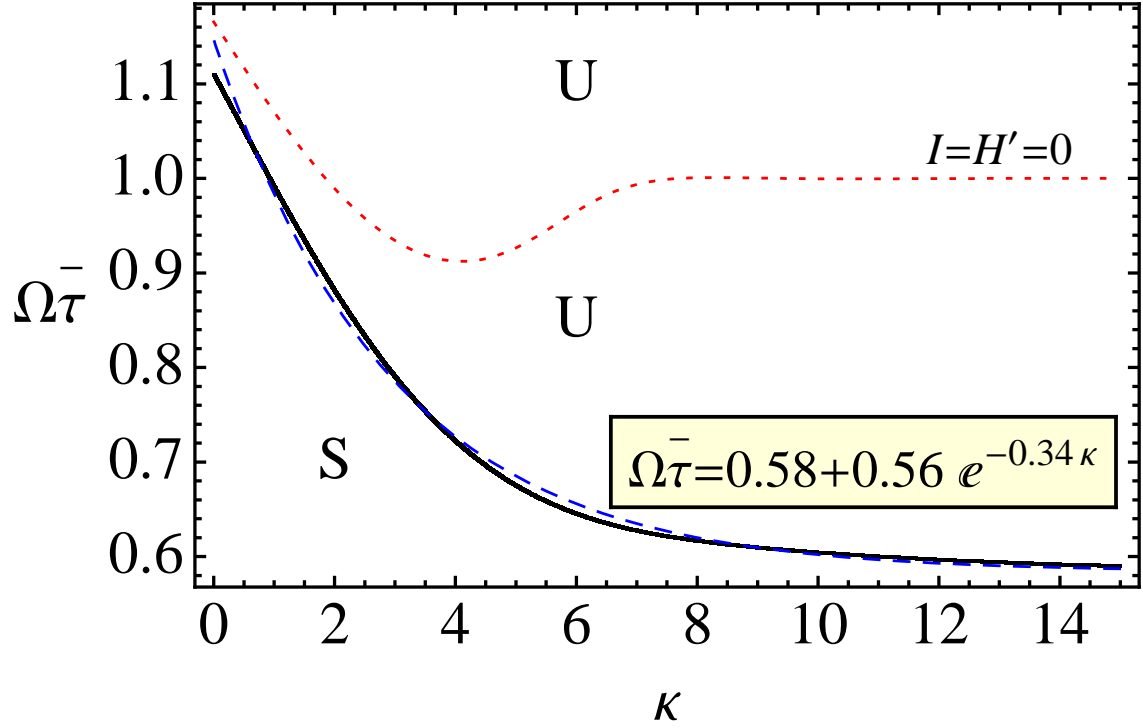}
\end{tabular}
\caption{ A numerical fit to the marginal stability curve gives an approximate scaling law in the form of an offset exponential relation between $ \Omega \bar{\tau}$ and $\kappa$. The marginal stability curve (solid, in black) of Fig.~{\ref{fig:fig4}} has been replotted along with the fitted curve (dashed, in blue).  The dotted curve (in red) is obtained from the condition $H'$ or $I=0$.}
\label{fig:fig5}
\end{figure}

The points where solid and dotted lines meet in the curves of Figs.~\ref{fig:fig1} and \ref{fig:fig3} mark the marginal stability point for the respective $\kappa$ values. These  points are obtained for a range of $\kappa$ values and are plotted in ($\Omega \bar{\tau},\kappa $) space in Fig.~\ref{fig:fig4} by filled points.  They all lie on a single curve, which is analytically derived below.
Our numerical results further reveal that for the marginal stability points, the imaginary part of the eigenvalue of the mode is zero---in other words the mode loses stability through a saddle-node bifurcation.  It can easily be checked by inspection that $\lambda_I=0$ is one of the  solutions of Eq.~(\ref{lambda_I}) for any value of $\lambda_R$; 
however, it is not evident analytically that this is the only possible solution for $\lambda_R=0$, and our numerical results have helped us confirm that this is indeed the case.
Hence, putting $\lambda_R =\lambda_I =0$ in Eq.~(\ref{lambda_R}) we get the following integral relation between the parameters $\Omega, \tau_m$ and $\kappa$.
\begin{equation}
\int_{-1}^{1}G(|z|)\cos\left(  \Omega \tau_{m} |z|\right)\left[  1-\cos\left(\pi n z\right)\right]dz = 0
\label{marg_stab}
\end{equation}
Further, we have also observed that the most unstable perturbation is the one with the lowest mode number, namely $n=1$. Therefore Eq.~(\ref{marg_stab}) with $n=1$ defines the marginal stability curve, so the condition for synchronization takes the form
\begin{equation}  
   \int_{-1}^{1}G(|z|)\cos\left(  \Omega \tau_{m} |z|\right)\left[  1-\cos\left(\pi z\right)\right]dz > 0
\label{marg_stab2}
\end{equation}
The solid line in Fig.~\ref{fig:fig4} is the analytical curve of marginal stability defined by setting the left side of Eq.~(\ref{marg_stab2}) to zero, and it can be seen that the numerically calculated marginal values (represented by points) fit this curve perfectly. The figure also shows the stability curves obtained for the $n=2$ and $n=3$ perturbations 
(dashed and dotted lines, respectively) and these are seen to lie above the $n=1$ marginal stability
curve. We have carried out a numerical check for a whole range of higher $n$ numbers and the results are consistent with the above findings. 

For a system with constant delay $\tau$ (i.e. if $\tau_m|z|$ is replaced by $\tau$ in Eq.~(\ref{phase})), the $\cos(\Omega\tau)$ term can be taken outside of the integral in (\ref{marg_stab2}) and the remaining integrand is nonnegative. Hence, the synchronization condition in this case becomes simply 
\begin{equation}  \label{strogatz-cond}
 \cos(\Omega\tau)>0.	
\end{equation}
This agrees with the results obtained previously  for constant-delay systems  \cite{yeung99,earl03}.Thus our result, as given by Eq.~(\ref{marg_stab}), generalizes the condition (\ref{strogatz-cond}) to systems with space-dependent delays, and shows  a nontrivial relation between the spatial connectivity and delays for the latter case. 


In order to gain some intuition into the complex interaction between connectivity and delays, we have obtained an approximate expression for the marginal stability curve by a numerical fitting procedure, yielding the relation 
\begin{equation}  \label{fit}
	\Omega \bar{\tau} < 0.58 +0.56 e^{-0.34 \kappa}
\end{equation}
for the stability of synchronous oscillations.
Here, the left side involves the temporal scales of the dynamics (namely, it is the average time delay normalized by the oscillation period of the synchronized solution) while the right hand side involves the spatial scales of connectivity. In this view, the synchronization condition is a balance between the temporal and spatial scales.
For high connectivity ($\kappa\to 0$), the system can tolerate higher average delays in maintaining synchrony, and the largest allowable delays decrease roughly exponentially as the spatial connectivity is decreased. In the same figure we have also plotted with dotted curve the approximate condition (\ref{nec_cond}), which is found to lie above  the marginal stability curve in the entire range of $\kappa$. The disparity between the two curves becomes 
particularly noticeable  at large values of $\kappa$.

\section{Conclusions and Discussion}
We have investigated the existence and stability of the synchronous solutions of a continuum of nonlocally coupled phase oscillators with distance-dependent time delays. Our model system is a generalization of the original Kuramoto model by the inclusion of naturally occurring propagation delays. The equilibrium synchronous solutions of this system are shown to differ significantly from those of similar earlier models that had introduced simplifications either in the coupling or in the nature of the time delay. The equilibrium solutions of the lowest branch are seen to exhibit frequency suppression as a function of the mean time delay. We have carried out a linear stability analysis of the synchronous solutions and obtained a comprehensive marginal stability curve in the parameter domain of the system. Our numerical results show that the synchronous states become unstable via a saddle-node bifurcation process and the most unstable perturbation corresponds to an $n=1$ (or kink type) spatial perturbation on the ring of oscillators. These findings allow us to define an analytic relation, given by Eq.~(\ref{marg_stab2}), as a condition for synchronization. We have also obtained approximate forms for the synchronization condition that provides a convenient means of assessing the stability of synchronous states. Our results indicate an intricate relation between synchronization and connectivity in spatially extended systems with time delays.

\section*{ACKNOWLEDGMENTS}
GCS would like to acknowledge the hospitality of the Max-Planck Institute for the Physics of Complex Systems, Dresden, Germany, where part of the work was carried out during his sabbatical at BioND Group headed by Dr. Thilo Gross. GCS also acknowledges his discussions with Dr. Luis G. Morelli at MPI-PKS.

\newpage
%
\end{document}